\documentclass[10pt,conference]{IEEEtran}
\usepackage{graphicx} %
\usepackage{xargs} 
\usepackage[pdftex,dvipsnames]{xcolor}
\usepackage[frozencache=true,cachedir=minted]{minted}
\usepackage{hyperref}

\usepackage[colorinlistoftodos,prependcaption,textsize=small]{todonotes}
\setuptodonotes{inline}

\newcommandx{\vlad}[2][1=]{\todo[linecolor=blue,backgroundcolor=blue!25,bordercolor=blue,#1]{Vlad: #2}}

\makeatletter
\let\@float@c@listing\@caption
\makeatother

\newcounter{objective}[section]
\ifCLASSOPTIONcompsoc
  \usepackage[caption=false,font=normalsize,labelfont=sf,textfont=sf]{subfig}
\else
  \usepackage[caption=false,font=footnotesize]{subfig}
\fi

\title{Improving Quantum Developer Experience with Kubernetes and Jupyter Notebooks}
\author{
\IEEEauthorblockN{Otso Kinanen}
\IEEEauthorblockA{
\textit{University of Jyväskylä}\\
Jyväskylä, Finland \\
otso.j.r.kinanen@jyu.fi}\\
\and
\IEEEauthorblockN{Andrés D. Muñoz-Moller}
\IEEEauthorblockA{
\textit{University of Jyväskylä}\\
Jyväskylä, Finland \\
andres.d.munozmoller@jyu.fi}\\
\and
\IEEEauthorblockN{Vlad Stirbu}
\IEEEauthorblockA{
\textit{University of Jyväskylä}\\
Jyväskylä, Finland \\
vlad.a.stirbu@jyu.fi}\\
\and
\IEEEauthorblockN{Tommi Mikkonen}
\IEEEauthorblockA{
\textit{University of Jyväskylä}\\
Jyväskylä, Finland \\
tommi.j.mikkonen@jyu.fi}
}
\date{January 2024}

\begin{document}

\maketitle

\begin{abstract}
Quantum computing proposes a revolutionary paradigm that can radically transform numerous scientific and industrial application domains. To realize this promise, new capabilities need software solutions that are able to effectively harness its power. However, developers face significant challenges when developing quantum software due to the high computational demands of simulating quantum computers on classical systems. In this paper, we investigate the potential of using an accessible and cost-efficient manner remote computational capabilities to improve the experience of quantum software developers. %
\end{abstract}


\section{Introduction}

Quantum computing holds great promise as a revolutionary technology that can transform various scientific and industry fields. By harnessing the principles of quantum mechanics, quantum computers can perform complex calculations and solve problems that are currently intractable for classical computers. This promises breakthroughs in areas such as cryptography, optimization, drug discovery, materials science, or machine learning. %
Although quantum advantage has been declared in experiments where quantum computing hardware has shown to provide a significant computational advantage over classical alternatives in specific problems~\cite{sandersquantum}, we still have to work for the foreseeable future with Noisy Intermediate-Scale Quantum (NISQ) computers. These computers employ a hybrid computational model in which a classical computer controls a noisy quantum device. %
Even as NISQ devices are not capable of providing the quantum advantage promised by quantum algorithms~\cite{Chen2023}, they are an invaluable platform for research and experimentation.

However, even with the steady advancements in terms of qubit counts~\cite{preskill2023quantum, gill2022quantum}, current NISQ computers are out of reach of most developers due to their scarcity and high operational costs. Therefore, quantum software developers have to rely on simulators running on classical computers to experiment with quantum software. While it is straightforward to start the development process on commonly used hardware, running larger circuits necessitates specialized graphical processing units (GPU) that are found in high-end consumer products (e.g. mobile workstations) or high-performance computing infrastructure (e.g. clusters of GPUs). Therefore, a developer would require either deep technical knowledge to configure the software stack required for using the advanced GPU capabilities to be able to run circuits up to 31 qubits~\cite{faj2023benchmark}, or access to a supercomputer for running circuits with 40 qubits~\cite{WILLSCH2022108411}.

Our approach to improve the quantum software development experience is to execute the quantum software routines on Qubernetes clusters \cite{qubernetes}, which manage the necessary computational resources. The solution is packaged as an easy-to-use Jupyter kernel, therefore the developers are not directly exposed to the complexities of operating the cluster where the quantum routines are executed, allowing them to switch between the local and the remote development environments when the number of qubits in the quantum circuits become large.

The rest of the paper is organised as follows. Section~\ref{sec:background} presents the background and motivation behind this work. Section~\ref{sec:solution} describes the implementation of the solution. Section~\ref{sec:evaluation} describes the test environments and discusses the performance. Concluding remarks and future work are presented in Section~\ref{sec:conclusion}.

\section{Background and motivation}
\label{sec:background}

\begin{figure*}
    \centering
    \includegraphics[width=0.8\textwidth]{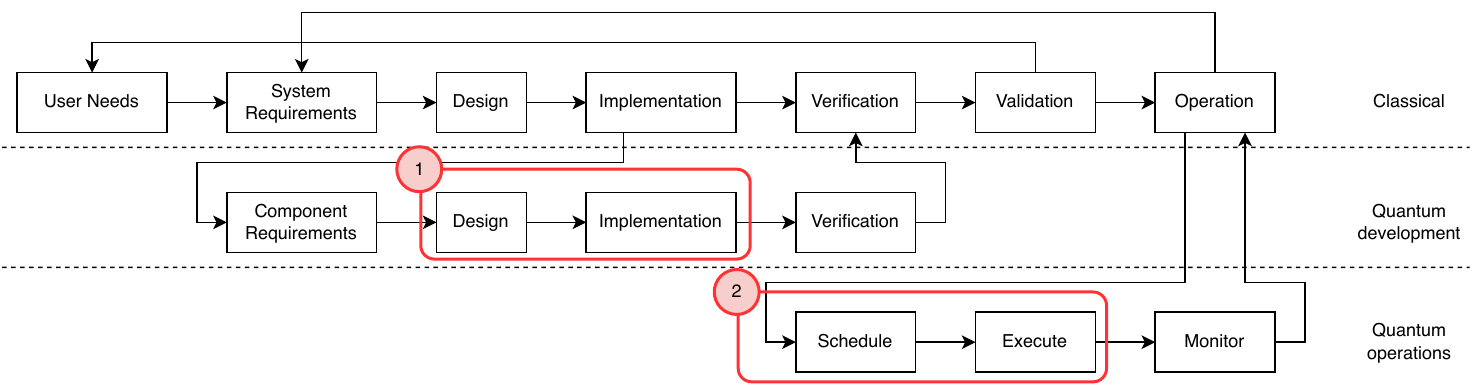}
    \caption{Quantum software development life cycle and areas where developers and operators interact with quantum hardware or simulators: (1) the design and development of quantum algorithms, and (2) scheduling and executing the computation on an available and capable quantum computer}
    \label{fig:sdlc}
\end{figure*}

\subsection{Software development life cycle}

The software development life cycle (SDLC) of hybrid classic-quantum applications consists of a multi-faceted approach \cite{stirbu2023fullstack}, as depicted in Figure~\ref{fig:sdlc}. At the top level, the classical software development process starts by identifying user needs and deriving them into system requirements. These requirements are transformed into a design and implemented. The result is verified against the requirements and validated against user needs. Once the software system enters the operational phase, any detected anomalies are used to identify potential new system requirements, if necessary. A dedicated track for quantum components is followed within the SDLC~\cite{sdlc}, specific to the implementation of quantum technology. The requirements for these components are converted into a design, which is subsequently implemented on classic computers, verified on simulators or real quantum hardware, and integrated into the larger software system. During the operational phase, the quantum software components are executed on actual quantum hardware. The scheduling ensures efficient utilization of the scarce quantum hardware resources, while monitoring capabilities enable the detection of anomalies throughout the operational stage.

As quantum computers are a scarce resource, it is not practical to develop quantum software components directly on hardware. Instead, developers should use simulators that use commonly available and less expensive classical resources (e.g., CPUs and GPUs~\cite{GUTIERREZ2010283}), for the early stages of development and testing. %
Later on, they can use more sophisticated simulators that are able to simulate the noise of actual hardware. Only when the components are mature enough the development can be continued on quantum processing units~(QPU), the actual hardware that will be used during the execution phase. However, as the implementation of quantum software stack trades off the visibility of the execution process for usability \cite{stirbu2024interplay}, developers have to experiment and iterate on devices and simulators to determine the actual behaviour of their programs. This approach ensures that the use of quantum resources is efficient and effective.

\subsection{Quantum development toolkits and simulators}

\begin{figure}[t]
    \centering
    \includegraphics[width=0.6\columnwidth]{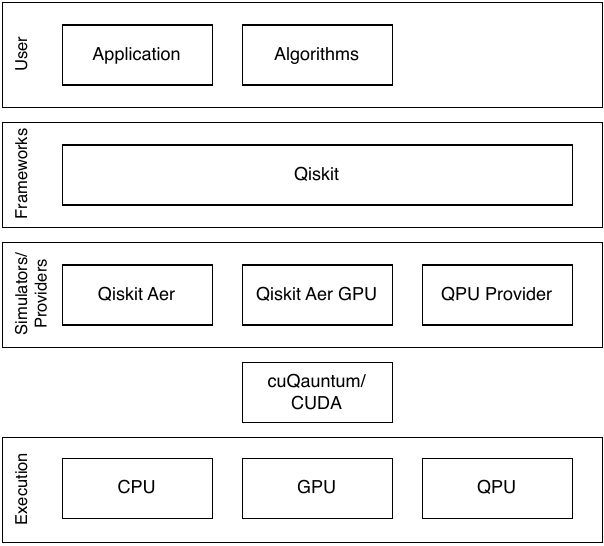}
    \caption{A layered view at the Qiskit software stack}
    \label{fig:executionstack}
\end{figure}

Qiskit is a Python library and a quantum development toolkit designed to accommodate different types of Quantum Computers in NISQ era. It allows algorithm designers develop applications leveraging quantum computing, circuit designers to optimize cirquits and explore its properties like error correction, verification and validation. Qiskit offers also tools to research and optimize gates, with precise control and ability to explore noise, apply dynamical decoupling and perform optimized control theory. Qiskit is an open-source project and currently offers dozens of additional libraries, plugins, simulator backends, application packages for multiple domains such as machine learning, physics, chemistry and finance and other related projects available. In Qiskit there are also several transpiler plugins available for users to optimize and interact with the transpiling process\footnote{\href{https://qiskit.github.io/ecosystem/}{https://qiskit.github.io/ecosystem/}}. Among Qiskit, IBM offers OpenQASM and OpenPulse. OpenQASM is an imperative language whose main purpose is to act as an intermediate representation for high-level compilers for QC hardware. It offers precise control over gates, measurement and conditionals\footnote{\href{https://openqasm.com/intro.html}{https://openqasm.com/intro.html}}. OpenPulse is a specification for pulse-level control, for general-purpose QC and it is designed to be hardware architecture agnostic and enable experimentation with a higher level of control \cite{mckay2018qiskit}. Qiskit Aer\footnote{\href{https://qiskit.github.io/qiskit-aer/index.html}{https://qiskit.github.io/qiskit-aer/index.html}} is Qiskit library with high-performance QC simulators and noise models. Some simulators included in Aer have support for leveraging Nvidia CPUs with Cuda version 11.2 or newer. Qiskit, Qiskit  Aer and Cuda relations in the development and execution environment are presented in Figure \ref{fig:executionstack}

PennyLane\footnote{\href{https://pennylane.ai/}{https://pennylane.ai/}} is a Python library specialized in machine learning for quantum computing by enabling the use of popular, commonly used classical machine learning frameworks, like TensorFlow \footnote{\href{https://www.tensorflow.org/}{https://www.tensorflow.org/}}. PennyLane is designed to support various executions with variable QC simulators and actual QC hardware, handling the communication with the device and compiling the circuits. The library includes a basic simulator backend and has GPU enabling PennyLane Lightning\footnote{\href{https://docs.pennylane.ai/projects/lightning/}{https://docs.pennylane.ai/projects/lightning/}} plugin with three different high-performance backends. Lightning GPU uses NVIDIA cuQuantum SDK to enable accelerating the simulation of quantum state vectors, it supports CUDA-capable Nvidia GPUs. 

Nvidia CUDA\footnote{\href{https://developer.nvidia.com/cuda-zone}{https://developer.nvidia.com/cuda-zone}} is a computing platform developed for GPUs, for computationally demanding tasks suitable for parallel computing with up to thousands of threads. cuQuantum\footnote{\href{https://docs.nvidia.com/cuda/cuquantum/}{https://docs.nvidia.com/cuda/cuquantum/}} is an SDK based on CUDA, offering libraries for Quantum computing, with two libraries, cuStateVec for state vector computation and cuTensorNet for tensor network computation. cuStateVec is used by gate-based general quantum computer simulators, providing measurement, gate application, expectation value, sampler and state vector movement. CuStateVec library is available for Cuda versions 11 and 12. Nvidia cuQuantum is used by both PennyLane Lightning and Qiskit Aer, for their GPU-powered quantum simulator backends. 

Developing across all target execution environments exposes the quantum software developer to a wide range of technologies that force them to balance their primary development activities with deep dives into operational aspects like configuring and maintaining their development environments or getting access to compatible hardware accelerators for running the relevant simulators. For example, Fig. 2 provides an overview of the software stack that application or algorithm developers using the Qiskit tools must to be aware of. The situation is similar for other mainstream toolkits like PennyLane or Cirq\footnote{\href{https://quantumai.google/qsim/cirq\_interface}{https://quantumai.google/qsim/cirq\_interface}}. Experimental programming toolkits like Eclipse Qrips \cite{qrisp}, leverage the existing Cirq or Qiskit assets to be able to execute circuits on GPU-accelerated simulators.

\subsection{Notebooks}

JupyterLab\footnote{\href{https://jupyter.org}{https://jupyter.org}} offers a versatile and user-friendly interactive computing platform suitable for data science, scientific computing, machine learning, and quantum computing. With its flexible architecture and extensive plugin ecosystem, it allows its users to develop customized workflows tailored to their specific needs, such as data exploration, prototyping algorithms or creating interactive presentations.

The key enabler of Jupyter is the notebook, an interactive and collaborative document formed by a collection of cells that can contain code, Markdown\footnote{\href{https://spec.commonmark.org/current/}{https://spec.commonmark.org/current/}} formatted text, equations or interactive widgets. A kernel is a computational engine that executes the code contained within the notebook. Jupyter supports multiple programming languages through different kernels, such as Python, R, Julia, and others. Users can select the desired kernel depending on their preferred programming language for a specific notebook. These combined capabilities allow scientists and algorithm developers to perform their work using a combination of code, explanatory text, and visualizations, making it easier to experiment, iterate, and document the development process.

JupyterHub\footnote{\href{https://jupyter.org/hub}{https://jupyter.org/hub}} expands the functionality of JupyterLab to groups of users, giving them access to computational environments and resources without the burden of installation and maintenance tasks. The project provides two distributions: \textit{The Littlest JupyterHub} -- suitable for small group of users, typically less than 100, can be installed on a single virtual machine, and \textit{Zero to JupyterHub for Kubernetes}\footnote{\href{https://z2jh.jupyter.org/en/latest/index.html}{https://z2jh.jupyter.org/en/latest/index.html}} -- suitable for large number of user, makes extensive use of container technologies, cloud resources and infrastructure. The container that runs JupyterLab can be customised following the Jupyter Docker Stacks\footnote{\href{https://jupyter-docker-stacks.readthedocs.io/en/latest/index.html}{https://jupyter-docker-stacks.readthedocs.io/en/latest/index.html}} convention, allowing the user to run quantum algorithms in GPU accelerated simulators like Qiskit Aer or PennyLane Lightning. However, as the pod life cycle is linked to the user session, the GPU is locked by the user's pod regardless if the Python kernel executes code or not, a utilization pattern that is not optimal.

\begin{figure}
    \centering
    \includegraphics[width=0.7\columnwidth]{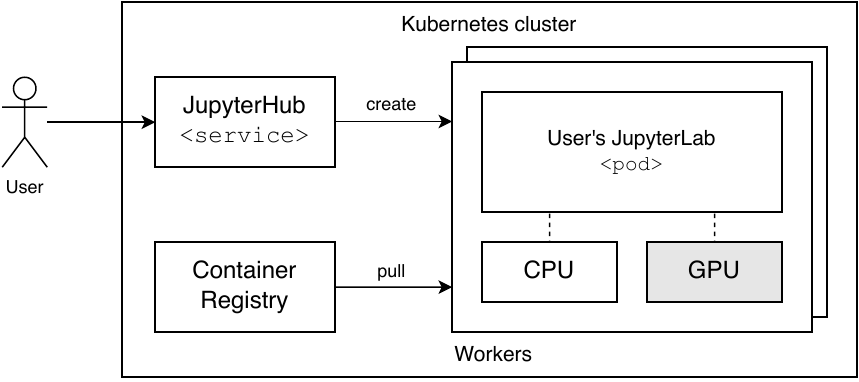}
    \caption{JupyterHub on Kubernetes architecture}
    \label{fig:jupyterhub}
\end{figure}

\subsection{Computing at-scale paradigms}%

Cloud computing allows the development of scalable applications~\cite{cloudnativeapps}, which rely on computing resources like computing power, storage and databases that are accessed on a pay-per-use basis. Through the extensive use of application programming interfaces (APIs), teams formed of software developers and operators can scale these resources up and down in response of the users' needs. This entails designing applications as small, loosely coupled components that can be bundled with their dependencies into portable containers and deployed on the immutable infrastructure. Furthermore, integrated monitoring and logging offer valuable insights into performance, health, and behaviour, empowering a swift response to potential anomalies. Kubernetes is the industry-standard container orchestration platform for automating deployment, scaling, and management of containerized cloud-native applications. Developed as an open-source solution by Cloud Native Computing Foundation (CNCF)\footnote{\href{https://www.cncf.io/}{https://www.cncf.io/}}, together with the myriad of projects that offers supporting functionality, it allows users to deploy applications on the managed infrastructure of the major cloud providers (e.g., AWS EKS\footnote{\href{https://aws.amazon.com/eks/}{https://aws.amazon.com/eks/}}, Azure AKS\footnote{\href{https://azure.microsoft.com/en-us/products/kubernetes-service}{https://azure.microsoft.com/en-us/products/kubernetes-service}}, or GCP GKE\footnote{\href{https://cloud.google.com/kubernetes-engine}{https://cloud.google.com/kubernetes-engine}}), smaller or regional cloud providers, or on-prem -- using own infrastructure.

Qubernetes \cite{qubernetes} (or Kubernetes for Quantum) exposes the quantum computation concepts like tasks and hardware capabilities following established cloud-native principles, as Kubernetes jobs and quantum-capable nodes. Following this conventions, allows the seamless integration of quantum computing into the larger Kubernetes ecosystem.

High-performance computing (HPC) relies on using supercomputers and parallel processing techniques to solve complex computational problems quickly and efficiently, in application domains that require massive computational power~\cite{sterling2017hpc}. HPC systems typically consist of multiple interconnected processors or nodes that work together to execute tasks in parallel, enabling large-scale simulations, data analysis, and scientific computations, leveraging the Open Message Passing Interface~(OpenMPI\footnote{\href{https://www.open-mpi.org}{https://www.open-mpi.org}}) compatible architectures.

Quantum computing enables the existing base of cloud-native and HPC applications to accelerate appropriate computational tasks. Two notable approaches for integrating the two software stacks are HPC-QC~\cite{hpc-qc-linking}, which uses the OpenMPI, and XACC~\cite{XACC} approach based on the OSGi\footnote{\href{https://www.osgi.org}{https://www.osgi.org}} architecture. Similarly, Qiskit's quantum-serverless~\cite{quantum-serverless} proposes a cloud-based approach for running hybrid classical-quantum programs. The proposed programming model, conforming to the RAY\footnote{\href{https://www.ray.io}{https://www.ray.io}} computing framework, makes it easy to scale Python workloads on a Kubernetes cluster in which the quantum execution environment is represented by a distributed Qiskit runtime that allows transparent access to multiple QPUs. Despite all these efforts, the integration of quantum computing into classical paradigms is fragmented. The EuroHPC aims to address this with the \textit{Universal Quantum Access}~\cite{eurohpc} development.

\section{The Jupyter kernel for Qubernetes}
\label{sec:solution}

\subsection{System architecture and components} 

\begin{figure}
    \centering
    \includegraphics[width=\columnwidth]{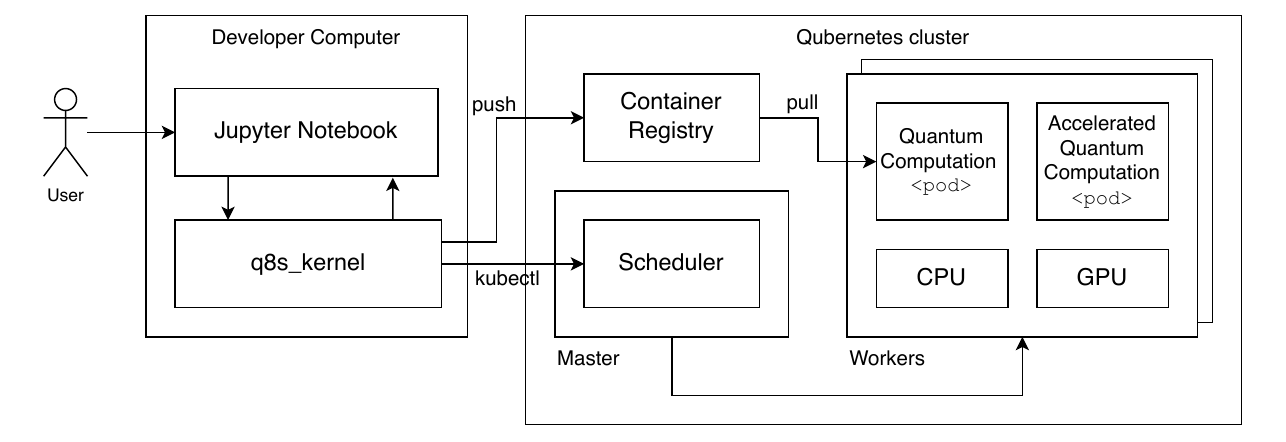}
    \caption{System architecture}
    \label{fig:system}
\end{figure}

The solution enables a quantum software developer to run quantum routines or programs using GPU-accelerated simulators (e.g. Qiskit Aer or Pennylane Lightning) on a Qubernetes cluster with better computational resources compared to their personal laptop. The solution involves a custom Jupyter kernel\footnote{\href{https://github.com/torqs-project/q8s-kernel}{https://github.com/torqs-project/q8s-kernel}} (e.g., \texttt{q8s\_kernel}), and a compatible cluster that has at least one quantum capable node that allows the execution of GPU-accelerated containers via the Nvidia Container Toolkit\footnote{\href{https://docs.nvidia.com/datacenter/cloud-native/container-toolkit/latest/index.html}{https://docs.nvidia.com/datacenter/cloud-native/container-toolkit/latest/index.html}}. To utilize the solution, the developer must install the kernel and specify the location of the configuration file of the cluster (e.g., \texttt{kubeconfig}\footnote{\href{https://kubernetes.io/docs/concepts/configuration/organize-cluster-access-kubeconfig/}{https://kubernetes.io/docs/concepts/configuration/organize-cluster-access-kubeconfig/}}) as an environment variable. Through the user interface of the Jupyter Notebook/Lab, the user can switch between the local development kernel (e.g. \texttt{IPython}) and the remote Qubernetes cluster. The system architecture and components are detailed in Fig.~\ref{fig:system}.

\subsection{Task execution model}

\begin{figure*}[!t]
    \centering
    \includegraphics[width=0.9\textwidth]{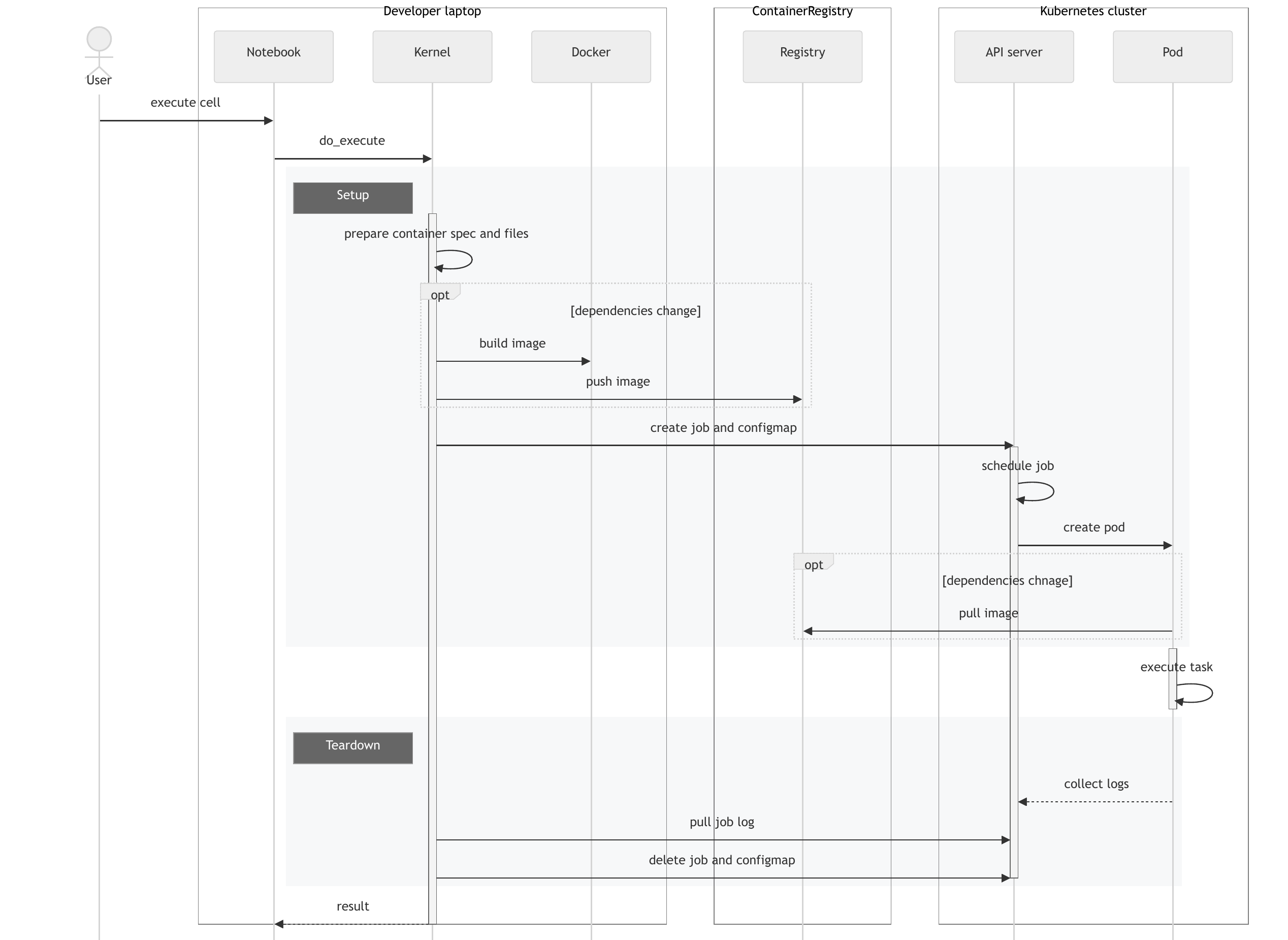}
    \caption{Executing a notebook cell on Kubernetes cluster using the q8s\_kernel}
    \label{fig:seq}
\end{figure*}

\begin{listing}[t]
\begin{minted}[fontsize=\scriptsize,xleftmargin=1em,linenos=true,highlightlines={12-13,16,19,23}]{yaml}
apiVersion: batch/v1
kind: Job
metadata:
  name: "quantum-job"
spec:
  template:
    metadata:
      name: "quantum-pod"
    spec:
      containers:
        - name: "quantum-task"
          image: registry.com/user/job-dependencies:v1
          command: ["python", "/app/main.py"]
          resources:
            requests:
              nvidia.com/gpu: '1' # requires GPU usage
          volumeMounts:
          - name: config-volume
            mountPath: /app
      volumes:
        - name: config-volume
          configMap:
            name: task-files #{"main.py": "code"}
      restartPolicy: Never
\end{minted}
\caption{Quantum job specification}
\label{listing:job}
\end{listing}

The execution flow is triggered by the user pressing the \textit{run} button in the notebook. When the kernel receives the \texttt{do\_execute} command, it detects the dependencies in the cell code and prepares the container specification (e.g., \texttt{Dockerfile} and \texttt{requirements.txt}), using as base a pre-build image that includes all dependencies for the CUDA version supported in the cluster. The kernel builds the image and pushes it to the container registry. Then it creates a Kubernetes Job specification that corresponds to the execution task (see Listing \ref{listing:job}), and a ConfigMap that contains the actual code that will be mounted as a volume in the Pod. Once the cluster API server receives the request, it schedules the job when the requested GPU resources are available. The Pod pulls the image from the Registry and executes the tasks. The kernel polls the API server for the Job's status waiting for completion, then collects the logs and cleans up, by deleting the Job and the ConfigMap. Depending on the container's exit code (e.g. success for 0, or failure otherwise), the kernel returns the result to the notebook on the \texttt{stdout} or \texttt{stderr} respectively. The kernel rebuilds the image and the pod pulls the image only when the dependencies change. The task execution sequence is depicted in Fig. \ref{fig:seq}.

\section{Discussion}
\label{sec:evaluation}

In this section, we introduce the test scenarios and evaluate the use of the custom Jupyter kernel from the \textit{ease of use} and \textit{cost effectiveness} perspectives.

\begin{table*}
    \caption{Hardware configurations used for the benchmark environment}
    \label{tab:hardware}
    \centering
    \begin{tabular}{lllll}
    \hline
    \bfseries Scenario & \bfseries Hardware category & \bfseries Model/Provider & \bfseries CPU & \bfseries GPU (CUDA compatible) \\
    \hline
         Baseline & Business laptop & Dell Latitude 7440 & Intel i5-1345U 16GB & - \\
         Mobile workstation & Mobile workstation & Dell Precision 7680 & Intel i9-13950HX 64GB & Nvidia GeForce RTX 4090 Laptop 16 GB\\
         Cloud GPU & Cloud server & puzl.cloud & 2 vCPUs up to 64GB & Nvidia A100 40GB\\
    \hline
    \end{tabular}
\end{table*}

We have tested the Jupyter kernel for Qubernetes in the following scenarios that we consider representative of how it will be used. The baseline consists of the user running the development environment (e.g. the Jupyter Notebook/Lab) and executing the quantum routine on his own laptop. The following test scenarios employ CUDA capable GPUs accessed remotely in Qubernetes clusters:

\textbf{Cluster with mobile workstation} - Users with better hardware share their computational resources (e.g. a mobile workstation) with the rest of the team. The users run the development environment similar to the baseline scenario, but the quantum routines are executed on the mobile workstation.

\textbf{Cluster with cloud GPUs} - The user runs the development environment on his own laptop and executes the quantum routine experiments on a Qubernetes cluster operated by a commercial entity. In our case, we have selected Puzl\footnote{\href{https://puzl.cloud/}{https://puzl.cloud/}}, a provider that offers access to Nvidia A100 40GB GPUs. The cost of using the GPU resources is approximately 1.6 EUR/h, in line with other cloud infrastructure providers. The charging model is based on effective utilization of the GPU resource, e.g., the effective time the Job runs to completion. The cluster is shared with the other Puzl users that execute their own workloads while our quantum routines are executed.

The detailed hardware configurations of the devices used in the test scenarios are described in Table~\ref{tab:hardware}.

\begin{figure}
    \centering
    \includegraphics[width=\columnwidth]{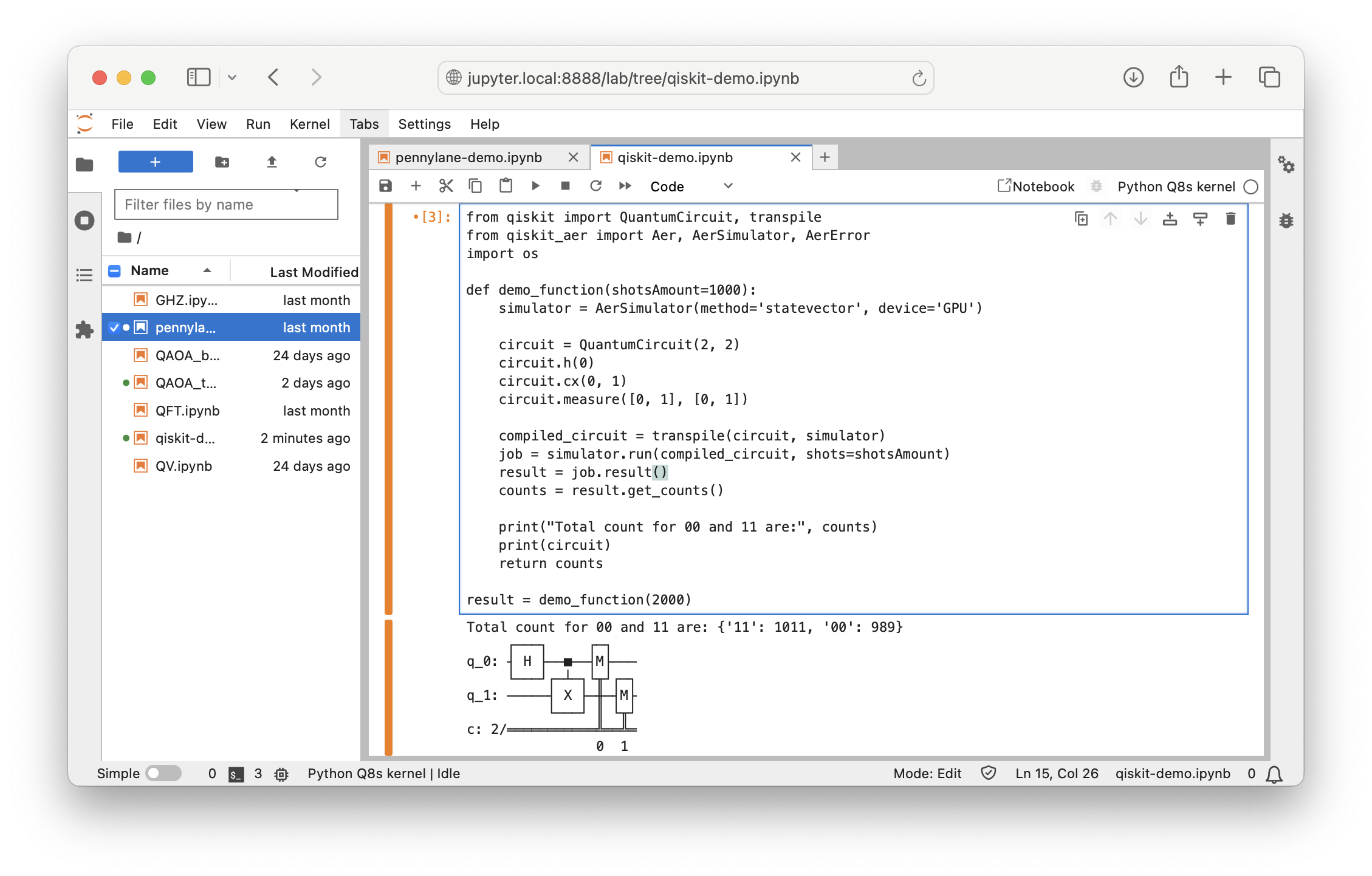}
    \caption{Jupyter lab environment configured for executing quantum computations on a remote cluster via the "Python Q8s kernel" (q8s\_kernel)}
    \label{fig:screenshot}
\end{figure}

\textbf{Ease of use} - The solution is perceived by the user as a standard Jupyter kernel, see Fig. \ref{fig:screenshot}. To function properly, the implementation relies on Docker and Kubernetes, widely used tools supported on a multitude of operating systems. The selection of the cluster where the quantum task execution is performed is achieved by providing the \texttt{kubeconfig} configuration file as an environment variable. The solution does not require a deep understanding of Kubernetes cluster management beyond the configuration file. As such the user is not exposed to the complexities of enabling access to the GPUs or configuring the computational layer of the CUDA, cuQauntum.

\textbf{Cost effectiveness} - The GPU resources available in the cluster are used only while the quantum accelerated job is executed. This behaviour allows the GPU resource of the mobile workstation to be used by other users of the cluster, maximizing its utilization of resources already acquired by the organization. Similarly, as the cloud GPU resource is charged per use, releasing the resource minimizes the cost. As such, the Jupyter kernel for Qubernetes increases the utilization in a cost efficient manner of the GPU resources, comparing with the standard JupyterHub on Kubernetes approach.

\section{Conclusion and future work}
\label{sec:conclusion}
This paper explores the potential of using remote computational resources available in Qubernetes clusters to enhance the experience of quantum software developers across three key aspects: execution speedups, ease of use and cost-effectiveness. To achieve this objective we have developed a custom Jupyter kernel that packages the notebook cells as Kubernetes jobs and executes them on clusters that have advanced CUDA computing capabilities. %
Moving forward, we plan to extend this functionality to other development environments beyond notebooks. %

\section*{Acknowledgements}

This work has been supported by the Academy of Finland (project DEQSE 349945) and Business Finland (project TORQS 8582/31/2022).

\end{document}